\documentclass[useAMS,usenatbib]{mn2e}

\usepackage{times}
\usepackage{graphicx}
\usepackage{rotating}

\def\pcm3{{\rm\thinspace cm^{-3}}}

\def\n_h{{\rm n_{H}}}

\def\NH1{{$N_{\rm HI}~$}}

\def\ga{{\rm\thinspace gauss}}

\def\approxlt{\mathrel{\hbox{\rlap{\lower .5ex \hbox {$\sim$}}
        \raise .15 ex \hbox{$<$}}}}
\def\approxgt{\mathrel{\hbox{\rlap{\lower .5ex \hbox {$\sim$}}
        \raise .15 ex \hbox{$>$}}}}

\def\la{\mathrel{\hbox{\rlap{\hbox{\lower4pt\hbox{$\sim$}}}\hbox{$<$}}}}
\def\ga{\mathrel{\hbox{\rlap{\hbox{\lower4pt\hbox{$\sim$}}}\hbox{$>$}}}}
\newbox\grsign \setbox\grsign=\hbox{$>$} \newdimen\grdimen
\grdimen=\ht\grsign
\newbox\simlessbox \newbox\simgreatbox \newbox\simpropbox
\setbox\simgreatbox=\hbox{\raise.5ex\hbox{$>$}\llap
     {\lower.5ex\hbox{$\sim$}}}\ht1=\grdimen\dp1=0pt
\setbox\simlessbox=\hbox{\raise.5ex\hbox{$<$}\llap
     {\lower.5ex\hbox{$\sim$}}}\ht2=\grdimen\dp2=0pt
\setbox\simpropbox=\hbox{\raise.5ex\hbox{$\propto$}\llap
     {\lower.5ex\hbox{$\sim$}}}\ht2=\grdimen\dp2=0pt

\title[Spectroscopically confirmed brown dwarfs]{Spectroscopically confirmed brown dwarf members of Coma Berenices and the Hyades}
\author[S. L. Casewell et al.]{S. L. Casewell$^{1}$\thanks{E-mail:
slc25@le.ac.uk},  S. P. Littlefair$^{2}$, M. R. Burleigh $^{1}$ \& M. Roy $^{1}$
\\ 
$^{1}$Department of Physics and Astronomy, University of Leicester, University Road, Leicester LE1 7RH, UK\\
$^{2}$Department of Physics and Astronomy, University of Sheffield, Sheffield
S3 7RH, UK\\
}
\begin{document}

\date{today }

\pagerange{\pageref{firstpage}--\pageref{lastpage}} \pubyear{2013}

\maketitle

\label{firstpage}

\begin{abstract}
We have obtained low and medium resolution spectra of 9 brown dwarf candidate
members of Coma Berenices and the Hyades using  SpEX on the NASA InfaRed Telescope Facility and LIRIS on the William Herschel
Telescope. We conclude that 7 of these objects are indeed late M or early L
dwarfs, and that two are likely members of Coma Berenices, and four of the
Hyades. Two objects, cbd40 and Hy3 are suggested to be a field L dwarfs,
although there is also a possibility that Hy3 is an unresolved binary belonging to the cluster. These objects have
masses between 71 and 53 M$_{\rm Jup}$, close to the hydrogen burning boundary
for these clusters, however only an optical detection of Lithium can confirm
if they are truly substellar.

\end{abstract}

\begin{keywords}
stars: low-mass, brown dwarfs, open clusters and associations:individual:Hyades,Melotte111
\end{keywords}

\section{Introduction}
The Hyades (Melotte 25; RA= 04 26.9, Dec= +15 52) and Coma Berenices (Melotte
111; RA = 12 23 00, Dec = +26 00 00, J2000.0) are the
closest open star clusters to the Sun at distances of 46.45$\pm$0.5 and 86.7$\pm$0.9 pc respectively
\citep{vanleuwen09}. Both these clusters are relatively mature at 625$\pm$50
Myr \citep{perryman98} and $\sim$500 Myr, and have low reddening values of
E(B-V)$<$1.0 and 3.2 mmag respectively
\citep{taylor06},  but this is where the similarities end. The Hyades has been
well studied (e.g. \citealt{reid92,gizis99,dobbie02}), and contains many
members ($\sim$400), whereas Coma Ber has been the
subject of relatively few detailed studies and membership is less certain. 

One of the reasons Coma Ber is often neglected is that it covers a
relatively large region on the sky ($\sim$100 deg$^{2}$), while being sparsely
populated.  A low proper motion (-11.5, -9.5 mas yr$^{-1}$) also makes kinematic surveys with a baseline of less than 10 years
difficult.
We performed a wide area search for new stellar members in
\citet{casewell06}, identifying 60 candidates. A more recent kinematic and photometric survey by \citet{kraus07} discovered 149 candidate members, of which
98 have a membership probability of $>$80 per cent, and determined that this
survey was complete to 90 per cent between the spectral types of F5 and M6.
A search for substellar members was also performed by \citet{casewell05} who
identified 13 new brown dwarf
candidates using optical and near-IR photometry. A similar optical photometric
survey was performed by \citet{melnikov12} who surveyed 22.5 deg $^{2}$ and
combined $RI$ photometry with 2MASS and UKIDSS. They discovered 12 new low
mass members down to a spectral type of M6-8.  \citet{terrien14} also
performed a photometric and proper motion survey of the cluster, discovering
8 new stellar members, and confirming the membership of 6 M
dwarfs discovered by \citet{kraus07} by measuring their radial velocity.
\citet{mermilliod08} also performed a radial velocity study of 69 solar type
stars, 46 of which were candidate cluster members, to search for close, low
mass companions. Of these 46, only 8 stars appear to be cluster members, and 6
additonal members were determined to be spectroscopic binaries, suggesting a
spectroscopic binary fraction of 22\%.

Coma Ber is only estimated to have 145$\pm$15 stars earlier than M6
\citep{kraus07} and a total mass of 112$\pm$16 M$_{\odot}$ compared to a
total mass of 300-460 M$_{\odot}$ for the Hyades \citep{reid92}. 

These mature clusters are expected to have undergone some
form of mass segregation as part of their dynamical evolution, with some
sources suggesting that as many as 50 per cent of the low mass members being lost
with time \citep{delafuente00}. In their large area study of Coma Ber
\citet{kraus07} suggest that some mass loss has occurred, but over a lower mass
range than their survey covered.  Despite there being clear evidence of mass
segregation in Praesepe, a similarly aged cluster, recent surveys have identified
many new candidate brown dwarfs \citep{baker10, boudreault12}, one of which has recently
been spectroscopically confirmed \citep{boudreault13}. There are also two
known T dwarf members of the Hyades \citep{bouvier08}, and \citet{goldman13}
reports the discovery of 43 new cluster members, many of which are low mass,
but are not in the substellar regime. They also suggest that one previously
known L0 dwarf, 2MASSIJ02330155+270406 \citep{cruz07}  is also a cluster member.

These open clusters provide excellent laboratories for studying brown dwarfs, mainly due to their coeval nature and known distances, 
ages and metallicities. Using brown dwarfs found in open clusters as benchmark objects is not new, as evidenced by the many field objects that are discovered, and latter associated with moving groups (e.g. \citealt{jameson08, casewell08}).  Recent work on the field population has separated brown dwarfs into gravity categories, using the suffixes $\alpha$, $\beta$, $\gamma$ and $\delta$ \citep{cruz09}. Where $\alpha$ is used to denote a field or "normal" gravity object, $\beta$ is used for an intermediate gravity object (~100 Myr), $\gamma$ a low gravity object (10-30 Myr) and  $\delta$ a very low gravity object (~1 Myr). At the age of the Hyades and Coma Ber however, the gravity is high enough to be very similar to, but slightly lower than that of field dwarfs  \citep{chabrier00}. This does, however, mean that any object with a lower than average gravity can be discounted as a potential cluster member. 

We recently performed proper motion and photometric searches of both
the Hyades and Coma Ber clusters to discover new substellar candidate members \citep{hogan08,casewell05}.
These surveys identified 12 new Hyades L dwarf candidates (using the moving
group method and near-IR photometry) and 13 new brown dwarf
candidates (using optical and near-IR photometry only) in Coma Ber.  These objects, if bona fide cluster members can be used as benchmarks
as their age and metallicity are known, and they provide a sample of brown dwarfs with a near, but lower than field gravity.    In this paper we present near-infrared
spectra of 9 of these candidates.

\section{Observations and  Data Reduction}
We observed two Coma Ber brown dwarf candidates using SpeX \citep{rayner03}
on the 3m InfaRed Telescope Facility (IRTF) and another two using LIRIS on the
William Herschel Telescope (WHT) on La Palma. All of the Hyades brown dwarf
candidates were observed using LIRIS and the WHT. These were, in general, the brightest of the identified candidate members.
\begin{table*}
\caption{\label{table1}Names, co-ordinates, near-IR magnitudes and proper motion for the
  observed objects. The top four objects (cbd) are Coma Ber candidates
  (magnitudes in MKO)and
  the remaining objects (Hy) are Hyades candidates (magnitudes from 2MASS).}
\begin{center}
\begin{tabular}{l c c c c c c c }
\hline
Name&RA&Dec& $J$& $H$&$K$&$\mu_{\alpha}$& $\mu_{\delta}$\\
&\multicolumn{2}{|c|}{J2000.0}&&&&\multicolumn{2}{|c|}{mas yr$^{-1}$}\\
\hline
cbd34&12:23:57.37&+24:53:29.0&15.94$\pm$0.2&-&14.93$\pm$0.2&-&-\\
cbd36&12:17:10.45&+24:36:07.6&16.28$\pm$0.2&-&15.11$\pm$0.2&-&-\\
cbd40&12:16:59.89&+27:20:05.5&16.30$\pm$0.2&-&15.14$\pm$0.2&-&-\\
cbd67&12:18:32.71&+27:37:31.3&17.68$\pm$0.2&-&16.10$\pm$0.2&-&-\\
\hline
Hy1&04:20:24.5&+23:56:13&14.6$\pm$0.03&13.85$\pm$0.04&13.42$\pm$0.04&148$\pm$7&-46.41$\pm$7\\
Hy2&03:52:46.3&+21:12:33&15.94$\pm$0.09&14.81$\pm$0.07&14.26$\pm$0.07&114.31$\pm$7&-36.95$\pm$7\\
Hy3&04:10:24.0&+14:59:10&15.75$\pm$0.07&14.78$\pm$0.07&14.17$\pm$0.06&102.46$\pm$7&-7.86$\pm$7\\
Hy4&04:42:18.6&+17:54:38&15.60$\pm$0.06&14.97$\pm$0.09&14.23$\pm$0.07&82.71$\pm$7&-21.25$\pm$7\\
Hy6&04:22:05.2&+13:58:47&15.50$\pm$0.06&14.81$\pm$0.06&14.25$\pm$0.08&99.37$\pm$7&-23.48$\pm$7\\
\hline
\end{tabular}
\end{center}
\end{table*}

\subsection{IRTF}
We observed cbd34 and cbd67 (\citealt{casewell05}; Table \ref{table1}) on the 14-04-2010 in thin cirrus and seeing of 0.5'' with SpeX using the low resolution prism
mode (R$\sim$200) and the 0.8'' slit. We used an ABBA nod pattern, and used 120s exposures
and 5 nods for cbd34 (600s) and 180s exposures and 8 nods for cbd67
(1440s). The spectra typically have S/N = 40.
The data were reduced and telluric corrected using \textsc{Spextool}
\citep{cushing04,vacca03}. 

\subsection{WHT}
We observed cbd36 and cbd40 (\citealt{casewell05}; Table \ref{table1}) on 18-03-2006 with LIRIS on the WHT.  We used the
ZJ grating and the 1'' slit with 300s second exposures and an ABBA nod pattern.
cbd36  was observed for 8 exposures (2400s) and cbd40 for 6 exposures (1800s). The seeing was
.8''-1'' and the S/N was 15 for both objects. 

The Hyades brown dwarf candidate members Hy1, Hy2, Hy3, Hy4 and Hy6
(\citealt{hogan08}; Table \ref{table1}) were
observed again using LIRIS with the 1'' slit and the ZJ grating (R$\sim$700) on the
10-12-2009.  The exposure times were 300s and an ABBA nod pattern was used. Hy1
was observed for 8 nods (2400s), Hy2 for 13 (3900s), Hy3 for 19 (5700s), Hy4
for 12 (3600s) and Hy6 for 8 (2400s). The seeing was $\sim$1-2'' and the S/N
was 15, 12, 23, 14, and 12 respectively.

The LIRIS data were all reduced using the \textsc{starlink} packages
\textsc{figaro} and \textsc{kappa}.

\section{Results}

We compared our reduced SpeX spectra to near-infrared standards from M8 to L4
(VB10: \citealt{burgasser04}, LHS2924: \citealt{burgasser06},  DENIS-P
J220002.05-202832.98: \citealt{burgasser06}, 2MASSW J2130446-084520:
\citealt{kirkpatrick10}; Kelu-1: \citealt{burgasser07},  2MASSW
J1506544+132106: \citealt{burgasser07b},  2MASSJ21580457-1550098:
\citealt{kirkpatrick10}) from the SpeX prism
library hosted by Adam Burgasser (http://pono.ucsd.edu/$\sim$adam/browndwarfs/spexprism/links.html).  DENIS-PJ220002.5-202832.98 is not defined as a standard, but was
included to provide a template for the L0 spectral class. 

The LIRIS spectra were compared to medium resolution spectra of M and L dwarfs
hosted in the IRTF spectral library maintained by John
Rayner (http://irtfweb.ifa.hawaii.edu/$\sim$spex/IRTF\_Spectral\_Library/). From M8
to L3 these objects were: VB10, LHS2924, 2MASS J07464256+2000321AB, 2MASS
J14392836+1929149, Kelu-1, 2MASS J15065441+1321060, 2MASS J22244381-0158521
\citep{cushing05, rayner09}.  2MASS J07464256+2000321AB has a spectral type of
L0.5 not L0, as there were no L0 spectra available. 

To determine the goodness of fit for the SpeX data we normalised the spectra
to have a value of one in the 1.2-1.3 $\mu$m region as in  \citet{burgasser10}, cross-correlated the data and
model to take into account any velocity shift or error in wavelength
calibration,
 and rebinned the spectra onto the model template wavelength range.
As the templates were also SpeX prism data no broadening was required.

We then computed the root
mean square (rms) of the residuals to the fit, normalised to the number of
points in the spectrum. The same process was used for
the LIRIS data, but the normalisation region was moved to 1.25-1.3 $\mu$m to avoid
the K I lines.  In this case, the templates were rebinned to data wavelength scale. The SpeX medium resolution template spectra have R$\sim$2000, compared to the LIRIS data which has R$\sim$700.  The results are shown in table \ref{rms} and Figures \ref{irtf}
for the SpeX data and \ref{lirishy} for the LIRIS data. 

In both sets of data no radial velocity shift has been taken into account, however, the radial velocity of Coma Ber is -1.2 kms$^{-1}$ \citep{vanleuwen09} and the Hyades is 40  kms$^{-1}$ \citep{perryman98}. Such velocities are too small to be determined using the resolution of both SpeX and LIRIS.

\begin{figure}
\begin{center}
\scalebox{0.35}{\includegraphics[angle=270]{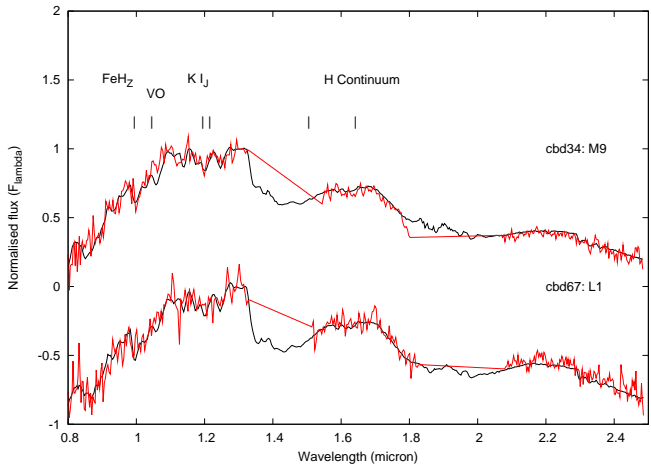}}
\caption{\label{irtf}The IRTF spectra (grey line) for cbd67 and cbd34 and their best fit templates (red line).
}
\end{center}
\end{figure}
\begin{figure}
\begin{center}
\scalebox{0.35}{\includegraphics[angle=270]{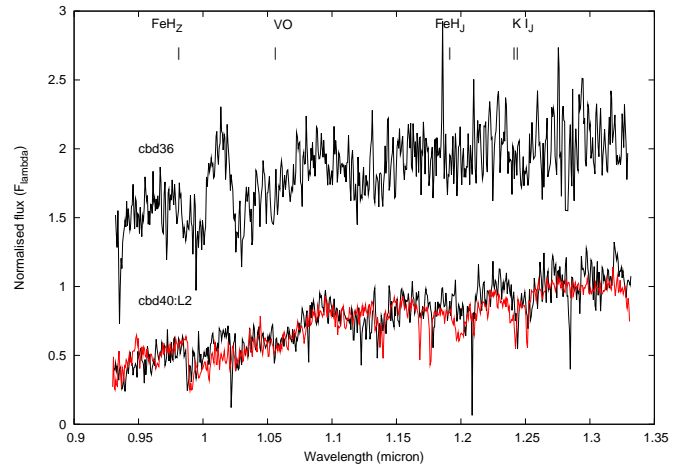}}
\caption{\label{irtf}The LIRIS spectra (grey line) for cbd36 and cbd40 and their best fit
  templates (red line). cbd36 is not a brown dwarf and therefore has no
  template matched to it.
}
\end{center}
\end{figure}

It was clear from the data on Hy2 that it is not a brown dwarf. It is missing
the FeH feature in the $Z$ band and the K I doublets in the $J$ band. It is
shown in Figure \ref{lirishy} for completeness, but without the template
spectra comparison. Similarly, cbd36 is of low S/N but also appears to be
missing the alkali lines and FeH feature. 
\begin{figure}
\begin{center}
\scalebox{0.35}{\includegraphics[angle=270]{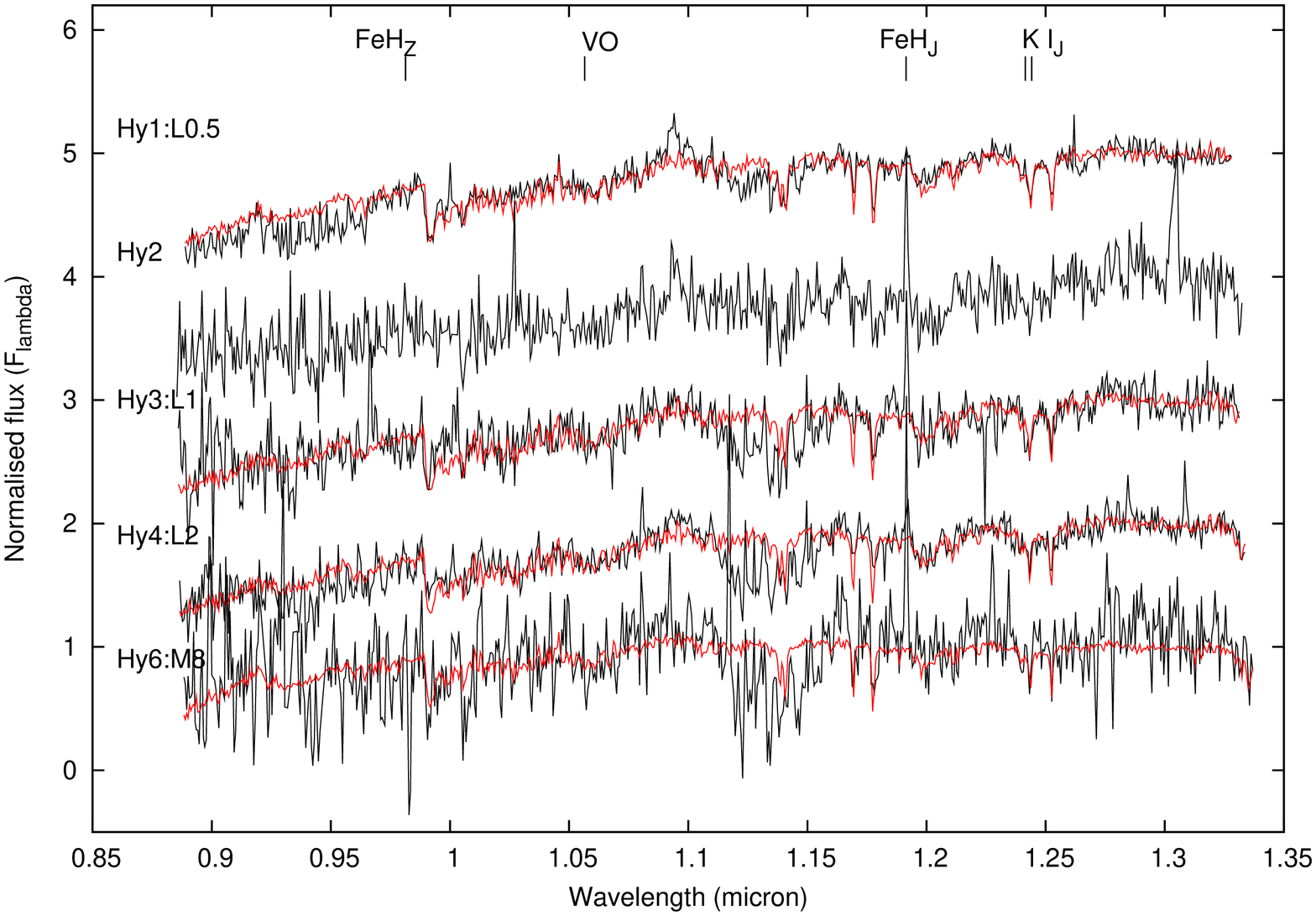}}
\caption{\label{lirishy}The LIRIS spectra of the Hyades brown dwarf candidates (grey line) and
  their best fit templates (red line). Hy2 is not a brown dwarf.
}
\end{center}
\end{figure}

\begin{table*}
\caption{\label{rms}Measured rms fits for the Hyades and Coma Ber brown
  dwarf candidates to the spectral templates. As all the spectra have been
  normalised to 1, the rms of the fit is also given.}
\begin{center}
\begin{tabular}{l c c c c c c }
\hline
Name&M8&M9&L0.5&L1&L2&L3\\
\hline
cbd34&0.081 &0.052 &0.068  &0.070 &0.142
&0.142 \\

cbd40&0.248 &0.160&0.138&0.136&0.126
&0.136\\
cbd67&0.136&0.093&0.080&0.071&0.092
&0.106\\
\hline
Hy1&0.208&0.124&0.113&0.117&0.140
&0.160\\

Hy3&0.240&0.195&0.189&0.187&0.189& 0.202\\
Hy4&0.229&0.163&0.152&0.149& 0.159&0.174\\
Hy6&0.310&0.335&0.355&0.361&0.368
&0.387\\
\hline
\end{tabular}
\end{center}
\end{table*}
From Figure \ref{irtf} and table \ref{rms} it can be seen that cbd67 has a
best fit to the L1 template, where cbd34 is best fit by an M9. cbd40 is
best fit by an L2 template, which is anomalous as 
 it is brighter than cbd67, and so should be of a later spectral type if they
 are both cluster members \citep{casewell05}.

For Hy1 the L0.5 template has the lowest rms, and represents
the FeH feature at 0.99 $\mu$m well, but the KI and Na I lines are not as deep in Hy1
as in the template, suggesting a spectral type of L0.

For Hy3 the L1 template has the best fit and replicates the shape and the line profiles of the spectra
well.

Hy4 is best fit by the L1 template. Hy6 has the  lowest
rms and $\sigma^2$ for the M8 template, but the spectrum has a low S/N  and
very little FeH edge in the $Z$ band, though the alkali lines are present in the
spectrum. 

\subsection{Spectral indices}
We have determined the spectral indices defined in \citet{allers13} for
FeH$_{Z}$ (0.99$\mu$m), VO$_{Z}$ (1.06$\mu$m), FeH$_{J}$ (1.20$\mu$m) and
KI$_{J}$ (1.244 and 1.253 $\mu$m).  For the SpeX data we also
determined the $H$-cont index. We were unable to do this for the LIRIS data as
it was taken in the ZJ filter and did not extend into the $H$ band. Similarly,
the FeH$_{J}$ was only determined for the higher resolution LIRIS data. The
indices are shown in table \ref{table2}. 

\subsubsection*{FeH$_{Z}$}
cbd34 and Hy1  have an FeH$_{Z}$ index consistent with the field population
as seen in \citet{allers13}.
 cbd40, cbd67 and Hy3 appear to have a much lower gravity score, as do Hy4 and Hy6,
 although it should be noted that Hy6 has a barely discernible FeH$_{Z}$
 feature in Figure \ref{lirishy}. 

\subsubsection*{VO$_{Z}$}
The VO$_{Z}$ index is only applicable to objects with spectral types later
than L0 \citep{allers13} so we have not included cbd34 in this analysis.  
All other objects except cbd67 have indices which are
consistent with their being field objects within the errors on this
index. cbd67 has a low value, but not within the low gravity range. If it were
of 0.5 spectral type later, it would be within the normal gravity range.

\subsubsection*{KI$_{J}$}
The KI$_{J}$ index was measured for all objects and all appear within the
field dwarf region. Hy6 appears to have a lower
value, which is not surprising as these lines are very weak. However, the large error bar on this measurement still puts it within the
field dwarf region.

\subsubsection*{H-cont}
Only cbd34 and cbd67 have an H-cont measurement as the LIRIS
grating does not cover the H band.  cbd34 has a slightly low gravity for its spectral type of M9, however, the
error bars on the measurement could move it into the field dwarf
sequence. cbd67 has a low value, for its spectral type, whereas most young
objects have a higher value.
 A
slightly lower gravity than the field measurement is consistent with it being a
member of Coma Ber.

\begin{table*}
\caption{\label{table2}Measured indices for the Hyades and Coma Ber brown
  dwarf candidates.}
\begin{center}
\begin{tabular}{l c c c c c}
\hline
Name&FeH$_{Z}$&VO$_{Z}$&FeH$_{J}$&KI$_{J}$&$H$-cont\\
\hline
cbd34&1.27$\pm$0.06&-&0.88$\pm$0.02&1.16$\pm$0.06&0.96$\pm$0.02\\
cbd40&1.09$\pm$0.06&1.11$\pm$0.04&0.92$\pm$0.17&1.14$\pm$0.03&-\\
cbd67&1.02$\pm$0.16&0.96$\pm$0.03&-&1.09$\pm$0.01&0.86$\pm$0.03\\
\hline
Hy1&1.20$\pm$0.04&1.14$\pm$0.03&1.18$\pm$0.030&1.12$\pm$0.03&-\\
Hy3&1.00$\pm$0.06&1.14$\pm$0.05&1.37$\pm$0.30&1.10$\pm$0.05&-\\
Hy4&1.17$\pm$0.05&1.13$\pm$0.05&1.31$\pm$0.13&1.16$\pm$0.03&-\\
Hy6&1.14$\pm$0.12&1.18$\pm$0.07&1.75$\pm$0.59&1.07$\pm$0.08&-\\
\hline
\end{tabular}
\end{center}
\end{table*}

\subsection{Equivalent Width measurements}

For the Hyades objects and the higher resolution data from LIRIS we were also able to measure the
equivalent widths (EWs) of the Na I lines at 1.1396$\mu$m and the K I lines at
1.1692, 1.1778, 1.2437 and 1.12529 $\mu$m. For this we fitted the
continuum in a window around the line, and Gaussians to the lines themselves
(see table \ref{table3}). Hy4 and Hy6 (Figure \ref{lirishy}) are of very low
signal to noise and so their measurements have large errors. The 1.1396 $\mu$m
feature is particularly unreliable for these objects as it is a blended
feature in the majority of these spectra.

We were unable to measure the EWs of the Coma Ber objects as the data is of
too low S/N.
\begin{table*}
\caption{\label{table3}Measured EWs in nm for the Hyades  brown
  dwarf candidates. The 1.2436 $\mu$m line is blended with a FeH feature and
  so we do not use the EW of this line to determine spectral type.}
\begin{center}
\begin{tabular}{l c c c c c}
\hline
Name&1.1396$\mu$m&1.1692$\mu$m& 1.1778$\mu$m& 1.2437$\mu$m&1.1252$\mu$m\\
\hline
Hy1&1.286$\pm$0.079&0.552$\pm$0.057&1.055$\pm$0.067&0.997$\pm$0.076&0.917$\pm$0.079\\
Hy3&2.345$\pm$0.130&0.385$\pm$0.082&0.906$\pm$0.094&0.639$\pm$0.093&0.449$\pm$0.099\\
Hy4&1.528$\pm$0.174&0.971$\pm$0.134&1.043$\pm$0.134&0.470$\pm$0.116&0.639$\pm$0.120\\
Hy6&1.613$\pm$0.417&1.153$\pm$0.224&0.760$\pm$0.201&0.744$\pm$0.321&0.812$\pm$0.326\\
\hline
\end{tabular}
\end{center}
\end{table*}

These EWs were compared to the EWs presented in \citet{cushing05}, and we used
the EW vs spectral type in table 8 and figure 17 to estimate a spectral type for these
objects. 
The EWs give a spectral type of between M9 and L1 for Hy1,  M8 for Hy3, L3
for Hy4 and L3 for Hy6. The EW of the 1.1692$\mu$m line for Hy3 suggests a
spectral type of M6, the 1.1252$\mu$m line M8 and the 1.1778$\mu$m line
L2. However, the earlier spectral types are inconsistent with the best
spectral fit (L1).

When compared to the EWs presented in
\citet{allers13} and using the gravity scores from the same work
($<$0.5=field dwarf gravity, 1=intermediate gravity, $>$1.5=very low gravity), 
the K I line measurements at 1.1252 $\mu$m show scores of 0
for Hy2 and Hy4, and 1 for Hy3 and Hy6. Hy1 has an anomalously high value. For
the K I 1.169$\mu$m line,  Hy3 has a low gravity score of 1, but the error bars
overlap the field dwarf region. All the other objects have a gravity score of
0 in line with field dwarfs.  For the K I 1.177 $\mu$m line, all
objects are consistent with having field gravity apart from Hy6, which scores a
1. It should be noted, however, that the large error bars on the measurement overlap with the field
dwarf region. The Na I 1.138 $\mu$m doublet gives scores of 1 for all objects,
however, the resolution is not good in this region of the spectrum and the region is
being affected by the telluric correction.

\section{Discussion}

Comparing the EW measurements to the spectral types determined from the
spectral fitting we conclude that Hy1 has a spectral type of L0.5, Hy3 a
spectral type between M8 and L0.5 and Hy4 has a spectral type between
L2 and L3 (Table \ref{spt}). Hy6 has a very inconclusive result, having a best
fit spectral template of M8, yet EW measurements suggesting L2. This is likely
due to to the low S/N of the data.
cbd67 is assigned a spectral type of L1 and cbd34, M9. cbd40 has been assigned
a spectral type of L2, although if this is correct, it is unlikely to be a
member of Coma Ber, as cbd67 is much fainter (by a magnitude in $J$ and
$K$). Indeed, 
comparing the $J-K$ colour for these two objects to the spectral type vs
colour relations presented in \citet{knapp04}, we see that cbd67 with a $J-K$
colour of 1.58 should have a later spectral type (L2-L3) than cbd40 which has
a $J-K$ colour of only 1.16 (L0.5) if they are both members of Coma Ber.

\begin{figure}
\begin{center}
\scalebox{0.35}{\includegraphics[angle=270]{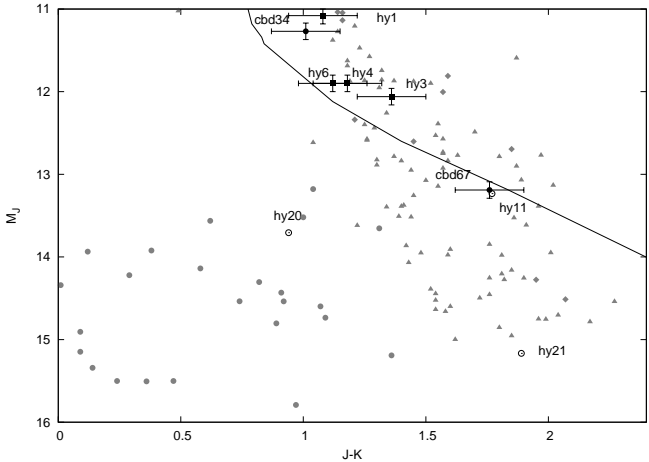}}
\caption{\label{jk}M$_{J}$, $J-K$ colour magnitude diagram for the Hyades
  candidates (filled squares) and Melotte candidates (filled circles). We have used the cluster distance (85
  pc) for the Coma Ber objects, and the kinematic distance given by
  \citet{hogan08} for the Hyads. The dusty \citep{chabrier00} isochrone for
  500 Myr is shown as the solid line. Field L dwarfs (grey filled triangles)
  and T dwarfs (grey filled circles) from
  \citet{faherty12} are also plotted as well as low gravity field L and M
  dwarfs (grey diamonds). The Hyades T dwarfs and Hy11 are plotted as open circles for comparison.
}
\end{center}
\end{figure}

The fit of the L0.5 standard to cbd40 is not a good fit. cbd40 has much weaker
alkali lines and FeH absorption than the template. As it is clear that cbd40
is a brown dwarf of likely spectral type L2, and has a gravity appropriate to field
objects, we conclude that it is likely to be a foreground field dwarf.
As can be seen in Figure \ref{jk} the remaining two objects have similar
colours to field dwarfs, and are consistent with the field L dwarf track and
the 500 Myr dusty model. Using the space density for M dwarfs and early L
dwarfs from \citet{cruz07} we estimate that our original survey \citep{casewell05} should
contain 1 field L dwarf and 2 field M dwarf contaminants. Our results so far
are consistent with this level of contamination. The Coma Ber objects were
not selected using proper motion, as there is not a sufficient epoch
difference between our Z band data and the UKIDSS Galactic Cluster Survey \citep{lawrence07}
data due to the low proper motion of Coma Ber. This is also compounded by
the sparsity of objects at this high Galactic latitude (b=38$^{\circ}$) making suitable
reference stars difficult to find.
 
Examining the spectral types and magnitudes of the Hyades objects, and using
the distance estimates from \citet{hogan08}, we see that Hy3 is the faintest
and reddest source which is not consistent with its spectral type when
compared to Hy4 (Figure \ref{jk}. Although, the photometric errors are large, and
will increase if the errors on the distance estimate are taken into account, it seems
unlikely that Hy3 could have an earlier spectral type than all the other
candidates and belong to the cluster. The fact that it is very red could
indicate 
that it is an unresolved binary as
it is close to the equal mass binary sequence of \citet{bannister07} in figure
4 of \citet{hogan08}, however no evidence is seen for this in the spectrum.  
Hy6  is the second brightest object and has an uncertain spectral type -
ranging between M8 and L2, probably due to the low S/N of the data.

In the Hyades, there are also two known T dwarfs, CFHT-Hy-20 (T2) and
CFHT-Hy-21 (T0) which have $J-K$ colours of 0.94 and 1.89 and $J$=17.02, and
$J$=18.48 respectively \citep{bouvier08}. One
of the 12 L dwarf candidates identified by \citet{hogan08}, Hy11 had
previously been identified as an L3 dwarf (2MASSW J0355419+225702) by
\citet{kirkpatrick99}. The colours of the Hyades candidates presented here are
consistent with the spectral types of these objects, and the spectral types of
the field objects from L1 to L3 span the gap between Hy4 and Hy6 and the L3
dwarf 2MASSW J0355419+225702 (Hy11).

\citet{hogan08} determined that of the 12 L dwarf candidates identified, two
were likely to be field dwarf contaminants. Of the five candidates we observed one (Hy2)
is not an L dwarf, and one object, Hy3, may be an equal mass binary, or
may be a field contaminant. This is broadly consistent with the estimated 
contamination rate.

\begin{table}
\caption{\label{spt}Spectral types for the Hyades and Coma Ber brown
  dwarf candidates.}
\begin{center}
\begin{tabular}{l c c}
\hline
Name&Spectral Type & Comment\\
\hline
cbd34&M9&\\
cbd36&-&Not a brown dwarf\\
cbd40&L2&Field dwarf\\
cbd67&L1&\\
\hline
Hy1&L0.5&\\
Hy2&-&Not a brown dwarf\\
Hy3&M8-L0.5& Possible field dwarf/cluster binary\\
Hy4&L2-L3&\\
Hy6&M8-L2&\\
\hline
\end{tabular}
\end{center}
\end{table}

If these M and L dwarfs are field objects they should have a log g of 5.40 for
their masses, however, if they are members of their respective clusters (age
$\sim$ 500 Myr) their gravity should be lower, at 5.24 \citep{chabrier00}. The
intermediate gravity ranking with a score of 1 is consistent with these
objects having a gravity of near to, but lower than the average field object.

Using the Dusty models of \citet{chabrier00}, their $K$ magnitudes, cluster
distance and $J-K$ colours  we have estimated the masses of
these brown dwarf candidates which can be seen in Table \ref{mass}. Owing to
the shape of the brown dwarf sequence, for objects with M$_{K}$ $>$ 11 the
temperature has been calculated by interpolating between the $J-K$ values
whereas for the brighter objects that all have similar $J-K$ colours we used
the $K$ magnitude.
All objects have masses between 71 and 53 M$_{\rm Jup}$, with Hy1 the most
massive at 71 M$_{Jup}$. cbd67 is the least massive, at 53
M$_{\rm Jup}$.  It is to be expected that the slightly younger Melotte
111 brown dwarfs are of lower mass for similar spectral types. The Dusty model
grid is quite coarse between 70 and 50 M$_{\rm Jup}$ and so even adding on an
error of 0.1 mag in both $J$ and $K$ does not produce a significant error in
mass.
\begin{table}
\caption{\label{mass}Masses and temperatures for the  Hyades and Coma Ber brown
  dwarf candidates. T$_{\rm eff}$ emp refers to the temperature derived from
  the empirical relationship in \citet{golimowski04} and T$_{\rm eff}$ model,
  refers to the temperature obtained from the Dusty models
  \citep{chabrier00}. Consequently if an object has a range of spectral types
  from our analysis there will be a range of temperatures derived from the
  empirical model.}
\begin{center}
\begin{tabular}{l c c c}
\hline
Name&Mass & T$_{\rm eff}$ emp&T$_{\rm eff}$ model\\
&M$_{\rm Jup}$&K&K\\
\hline
cbd34&64&2350&2145\\
cbd67&53&2200&1838\\
\hline
Hy1&71&2250&2310\\
Hy4&69&2000-2100&2289\\
Hy6&67&2500-2100&2230\\
\hline
\end{tabular}
\end{center}
\end{table}

We compared the temperature given by the effective temperature (T$_{\rm eff}$)
vs spectral type relationships in \citet{golimowski04} to those given by the dusty models for the
respective masses at 500 Myr. The Dusty models \citep{chabrier00} give T$_{\rm eff}$ of
2295 K for a 70 M$_{\rm Jup}$ object, 2048 K for a 60 M$_{\rm Jup}$ object and 1751 K
for a 50 M$_{\rm Jup}$ object. These temperatures interpolated to the masses
of the objects studied here can be seen in
Table \ref{mass}.

The temperature values derived from the empirical relations in \citet{golimowski04}  are slightly lower than those
provided by the Dusty models, but are broadly consistent. 
It should be noted here that these mass estimates are
consistent with there being a sequence of brown dwarfs in the Hyades, with the
two known T dwarfs CFHT-Hy-20 (T2) and CFHT-Hy-21 (T0) having estimated masses
of 50M$_{\rm Jup}$ \citep{bouvier08}.

These masses and effective temperatures indicate these objects are all very close to the H-burning limit. 
Therefore without an optical spectrum and a Lithium detection we are unable
to be sure whether these objects are truly brown dwarfs or stars.
However, the lack of low gravity scores for all of these objects indicate that
these objects are not young field objects, increasing the likelihood of their
being cluster members. These M and L dwarfs are also the remnants of the low mass star and brown dwarf population in these clusters, and while there are 
too few of them to make any firm conclusions about the mass functions of these clusters, they can be used as evidence of mass segregation and a a history of 
dynamical evolution that has ejected the low mass cluster members.

\section{Conclusions}
Using near-IR spectra from SpeX on IRTF and LIRIS on the WHT we have obtained
spectra of 9 brown dwarf candidate members of the Hyades and Coma Ber. We
have rejected cbd36 and Hy2 as brown dwarfs from their spectra. We also reject
cbd40 as a member of Coma Ber, and suggest that Hy3 may be a field object,
or an unresolved binary that is a cluster member.  The
remaining objects have spectral types ranging from M9 to L2 and masses
between 71 and 53 M$_{\rm Jup}$.
Using EWs and the indices defined by \citet{allers13} we have determined that
none of these objects have low gravity, which would indicate that they are
younger than the clusters they have been identified with, thus supporting
their cluster membership. 
These are the first spectroscopically confirmed L dwarfs in the Hyades cluster
and Coma Ber, however, optical spectra containing a lithium detection is
required to determine if these objects have a truly substellar nature.

\section{Acknowledgements}

The authors thank K. N. Allers for kindly providing the spectral indices tracks for field dwarfs that appear in Allers \& Liu, 2013.

SLC acknowledges support from the College of Science and Engineering at the
University of Leicester.

These results are based on observations made with the William Herschel
Telescope operated on the island of La Palma by the Isaac Newton Group in the
Spanish Observatorio del Roque de los Muchachos of the Instituto de
Astrofisica de Canarias. M Burleigh was a Visiting Astronomer at the Infrared Telescope Facility, which is operated by the University of Hawaii under Cooperative Agreement no. NNX-08AE38A with the National Aeronautics and Space Administration, Science Mission Directorate, Planetary Astronomy Program.

This research has made use of NASA's Astrophysics Data System Bibliographic
Services and has benefitted from the SpeX Prism Spectral Libraries, maintained by Adam Burgasser at http://pono.ucsd.edu/~adam/browndwarfs/spexprism

\bibliographystyle{mn2e}

\bibliography{hyades-1}
\label{lastpage}

\end{document}